\documentclass[twoside]{articlek}

\renewcommand{\refname}{REFERENCES}
\def\be{\begin{equation}}
\def\ee{\end{equation}}

\def\BibTeX{{\rm B\kern-.05em{\sc i\kern-.025em b}\kern-.08em
            T\kern-.1667em\lower.7ex\hbox{E}\kern-.125emX}}

\usepackage{graphics,epsfig}
\begin{document}
\sloppy
\twocolumn[{
\vspace*{1.7cm}   
\begin{center}
{\large\bf SIGNATURES OF STRONG ELECTROWEAK SYMMETRY
 BREAKING IN}
{\large\bf pp$\rightarrow$bbtt+X AT LHC.}\\

{\small M. Gintner, gintner@fyzika.uniza.sk, I. Melo, melo@fyzika.uniza.sk, and 
B. Trpi\v{s}ov\'a, trpisova@fyzika.uniza.sk

Katedra fyziky, \v{Z}ilinsk\'a univerzita, Univerzitn\'a 1, 01026 \v{Z}ilina, 
Slovak Republic}\\

\end{center}
\vspace*{1ex}

{\bf ABSTRACT.}We study the possible production of the  
$\rho$-resonance from the Strong Electroweak 
Symmetry Breaking sector at LHC. Due to enhanced coupling of $\rho$
to the top and bottom quarks we focus on  the process 
$\mathrm{pp} \rightarrow \mathrm{b}\bar{\mathrm{b}} 
\mathrm{t}\bar{\mathrm{t}}$ where either $\mathrm{b}\bar{\mathrm{b}}$ or
$\mathrm{t}\bar{\mathrm{t}}$ are the products of the $\rho$ decay.\\ 
}]

\section{INTRODUCTION}

Strong Electroweak Symmetry Breaking (Strong ESB or SESB)
 is an alternative mechanism of ESB, different from the Standard Model (SM) and 
Supersymmetry in that it generates
masses of elementary particles via new strongly interacting physics. 
SESB is motivated by the chiral symmetry breaking in QCD. Just as $\rho^{QCD}$
unitarizes $\pi \pi$ scattering in QCD, so does a
new strong vector (spin = isospin =1) resonance in the
form of an isospin triplet $\rho \;(\rho^{\pm}, \rho^0)$, with mass around 1~TeV scale, unitarize
WW scattering. The role of pions (Goldstone bosons) of the new strong interactions 
is played by the longitudinal W bosons to which the $\rho$ resonance is coupled strongly.
$\rho$ is thus a generic prediction of the SESB models \cite{Lane}. An effective Lagrangian 
description of the $\rho$ interactions with SM particles
was developed and has become known as the BESS model \cite{BESS}. 
This model is minimal in the sense that $\rho$ is the only new particle in the spectrum of SM
where it replaces the Higgs boson. 

We introduced modifications to the BESS model \cite{APS2001},\cite{APS2006}
which allow $\rho$ to couple significantly not only to the W bosons
but also to the top and bottom quarks. The study of the size of these couplings could 
indicate whether the mechanism of the W and Z mass generation is the same as 
the mechanism of
the top mass generation or not.
This motivates our study of the process(es)
pp$\rightarrow \rho$t$\bar{\mathrm{t}}$+X$\rightarrow$ 
b$\bar{\mathrm{b}}$t$\bar{\mathrm{t}}$+X 
and pp$\rightarrow \rho$b$\bar{\mathrm{b}}$+X$\rightarrow$ 
b$\bar{\mathrm{b}}$t$\bar{\mathrm{t}}$+X in this
contribution. (Previous studies found that pp$\rightarrow$jjW$^+$W$^-$+X is a 
good probe of $\rho$,
pp$\rightarrow$jjt$\bar{\mathrm{t}}$+X is overwhelmed by the 
pp$\rightarrow$t$\bar{\mathrm{t}}$+X background and pp$\rightarrow$t$\bar{\mathrm{t}}$W$^+$W$^-$+X
has a small cross-section).

\begin{figure} [h,t]                     
\begin{center}                        
\epsfig{file=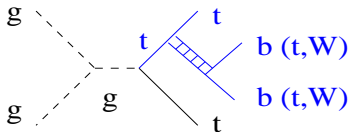,height=18mm,width=50mm,clip=,silent=,angle=0,
bbllx=0cm,bblly=0cm,bburx=6cm,bbury=2.5cm}  
\end{center}                         
\vspace{-2mm} \caption{Feynman diagram of  
pp$\rightarrow$b$\bar{\mathrm{b}}$t$\bar{\mathrm{t}}$+X.}
\end{figure}

\section{LAGRANGIAN}

The model is described by the effective chiral Lagrangian based on the
gauged non-linear sigma model respecting the symmetries 
of the Higgs sector of SM: SU(2)$_{\mathrm{L}}\times$U(1)$_{\mathrm{Y}}$ local and 
SU(2)$_{\mathrm{L}}\times$SU(2)$_{\mathrm{R}}$ global \cite{APS2001},\cite{APS2006}.
We will focus here on the interactions of the neutral $\rho^0$ with the 
top and bottom quarks which are relevant for pp$\rightarrow$t$\bar{\mathrm{t}}$
b$\bar{\mathrm{b}}$ (for completeness we also show $\rho^0$ interaction 
with longitudinal W bosons denoted as $\pi$ below):
\begin{eqnarray}
{\cal L} & = & 
+ \frac{b_2}{1+b_2}\frac{g_v}{2}\bar{t}_R \gamma^{\mu}t_R \rho_{\mu}^{0}+
\frac{b_1}{1+b_1}\frac{g_v}{2}\bar{t}_L \gamma^{\mu}t_L \rho_{\mu}^{0} 
\nonumber \\
& & 
 -\frac{b_2}{1+b_2}\frac{g_v}{2}\bar{b}_R \gamma^{\mu}b_R\rho_{\mu}^{0} 
- \frac{b_1}{1+b_1}\frac{g_v}{2}\bar{b}_L \gamma^{\mu}b_L\rho_{\mu}^{0} \nonumber \\
& & + i\frac{M_{\rho}^2}{2v^2g_v} 
\big( \pi^{-} \partial^{\mu} \pi^{+} - \pi^{+} \partial^{\mu} \pi^{-} \big)
\rho_{\mu}^{0},
\label{eq1}  
\end{eqnarray}
where $b_1, b_2, g_v$ are $\rho^0$ coupling constants, $v$ denotes the electroweak
scale 246~GeV and $M_{\rho}$ is the $\rho$ mass. 

On the basis of the BESS model studies we have for our model the low energy 
limit
$g_v \stackrel{>}{\sim} 10$. We do not have a strict low energy limit on $b_1$
and $b_2$. The unitarity limit requires that $g_v b_{1(2)}/4 <  2$.

\section{RESULTS}

For the pp$\rightarrow$t$\mathrm{\bar{t}}$b$\mathrm{\bar{b}}$ 
we took into account only the dominant gluon-gluon
channel with 155 Feynman diagrams. Unfortunately we
haven't been able to do the calculation for such a large number of diagrams.
Instead we singled-out 8 diagrams for 
pp$\rightarrow \rho$t$\bar{\mathrm{t}}$+X$\rightarrow$ 
b$\bar{\mathrm{b}}$t$\bar{\mathrm{t}}$+X ($\rho \rightarrow$ b$\bar{\mathrm{b}}$ signal), 8 diagrams for
pp$\rightarrow \rho$b$\bar{\mathrm{b}}$+X$\rightarrow$ 
b$\bar{\mathrm{b}}$t$\bar{\mathrm{t}}$+X ($\rho \rightarrow$ t$\bar{\mathrm{t}}$ signal) and treated the remaining diagrams as the
continuum background on the top of which we hope to see the $\rho$ signal. The number of background 
diagrams was further reduced by choosing the gauge invariant set of 35 diagrams
containing only quarks and gluons which dominate the rest. We call this set 
the ``QCD background".

The first step was to 
compute cross-sections  and generate events
using CompHEP \cite{comphep}. 
 The events were then passed to Pythia \cite{pythia} for decays and
hadronization and after that to Atlfast which simulates the ATLAS detector
effects. Everything was run within the Athena framework. 
The output from Atlfast was a ROOT file on which we did the final 
reconstruction. 

In Tab.1 we show cross-sections for several values of the coupling constants 
for both signal processes. As can be noticed, they range from tenths of fb up to 
a few hundred fb. Note the higher
values for the $\rho \rightarrow$ t$\bar{\mathrm{t}}$ signal as compared to
$\rho \rightarrow$ b$\bar{\mathrm{b}}$ by a factor of about 30. In Tab.1 we also
give the cross-section for the QCD 
background. As this number indicates, the QCD background is high but it 
dominates at small values of invariant masses m$_{\mathrm{tt}}$, 
m$_{\mathrm{bb}}$.

\begin{table}[ht]
\caption{Cross-sections of both signal processes for selected
values of the coupling constants $b_1$ and $g_v$ and the cross-section of the 
QCD background.}
\begin{tabular}{|l|c|c|r|r|}
\hline
\hbox{\hspace{1.cm}}&\hbox{\hspace{.4cm}}$b_1$ \hbox{\hspace{.3cm}} &
\hbox{\hspace{.4cm}}$g_v$ \hbox{\hspace{.3cm}} &
$\Gamma_{\rho}$[GeV]  & 
\hbox{\hspace{.4cm}}$\sigma$[fb]\hbox{\hspace{0.3cm}}\\
\hline \hline
& 0.02 & 20 & 4.1 & 0.2 \\
\cline{2-5}
$\rho \rightarrow \mathrm{b}\bar{\mathrm{b}}$& 0.08 & 20 & 42.4 & 4.3 \\
\cline{2-5}
& 0.08 & 40 & 166.4 & 17.8 \\
\hline
& 0.02 & 20 & 4.1 & 6.7 \\
\cline{2-5}
$\rho \rightarrow \mathrm{t}\bar{\mathrm{t}}$& 0.08 & 20 & 42.4 & 136.2 \\
\cline{2-5}
& 0.08 & 40 & 166.4 & 610.1\\
\hline
QCD & & & & 16388.8 \\
\hline
\end{tabular}
\end{table}

We chose  for the reconstruction the channel with one
charged lepton, $\mathrm{t}\bar{\mathrm{t}}\mathrm{b}\bar{\mathrm{b}} 
\rightarrow$b$\mathrm{W}^+ \bar{\mathrm{b}}\mathrm{W}^-$
b$\bar{\mathrm{b}}$$\rightarrow$bjj$\bar{\mathrm{b}}\mathrm{l}\nu_l$b$\bar{\mathrm{b}}$ (j denotes 
light jet, b stands for b-jet) which has the highest branching ratio among
all channels (43.5\%).  

The cuts imposed during the reconstruction were as follows: p$_{\mathrm{T}}$ 
of electron~$>$~30~GeV, p$_{\mathrm{T}}$ of muon~$>$~20~GeV, p$_{\mathrm{T}}$
 of jets~$>$~25~GeV. 
We assumed the b-tagging efficiency
of 50\%. One of the W bosons was reconstructed from the reconstructed lepton
and neutrino, the  remaining W was reconstructed from the light jets.
In the case of the leptonic decay the W mass (m$_{\mathrm{W}}$) constraint was 
used to determine the
longitudinal component of the neutrino.The jet-jet invariant mass
m$_{\mathrm{jj}}$ was required
to fall within the m$_{\mathrm{W}}\pm 25$~GeV window. For the reconstruction of 
each event 
we took such combination of two light jets and four b-jets (one can make 12
combinations to pair two W's and four b-quarks) which minimizes the 
expression

\begin{eqnarray}
\chi^2 & = & (m_{j_1j_2}-m_W)^2+(m_{W_1b_i}-m_t)^2+  \nonumber \\ & &  
(m_{W_2b_k}-m_t)^2,    
\end{eqnarray}

\noindent where  m$_{\mathrm{Wb}}$ is the W-b invariant mass and 
m$_{\mathrm{t}}$ is the mass of the top quark. The indeces $i,k=1,2,3,4$  
combine in such a way that for all combinations $i\neq k$ and the
order of the numbers in a combination matters.

In Fig.2 are shown reconstructed
distributions of the invariant mass $m_{\mathrm{bb}}$
of the b$\mathrm{\bar{b}}$ pair obtained for the 
$\rho \rightarrow$ b$\bar{\mathrm{b}}$
process (Fig.2a) and for the QCD background (Fig.2b). 
The resonance depicted in Fig.2a has mass $M_{\rho}$=1000~GeV
and width $\Gamma_{\rho}=42.4$~GeV. For both signal and background
we assumed integrated luminosity 100~fb$^{-1}$.
 $N$ is the total
number of events that passed through the cuts. 
Note the clear peak at 1000~GeV exhibited by the resonance. 

\section{CONCLUSIONS}

\noindent 
To compare signal and backround, we choose a window  $\pm$ 150 GeV around 
$m_{\mathrm{bb}}$ = 1 TeV. In this window we find 0.8 signal events 
(corresponding to the resonance in Fig.2a) and 8
background events (corresponding to the $\mathrm{m}_{\mathrm{bb}}$ distribution
in Fig.2b). The signal at this point seems weak and calls for the highest possible 
luminosity at LHC. However, further analysis is required which may lead to the improvement of the signal
to background ratio. The work on the reconstruction of the  $\rho\rightarrow$t$\mathrm{\bar{t}}$ signal which has a
higher cross-section is in progress.

\setlength{\unitlength}{1.0cm}
\begin{picture}(9,11.5)
\put(-0.4,6){\begin{picture}(9,5.5)
\epsfig{file=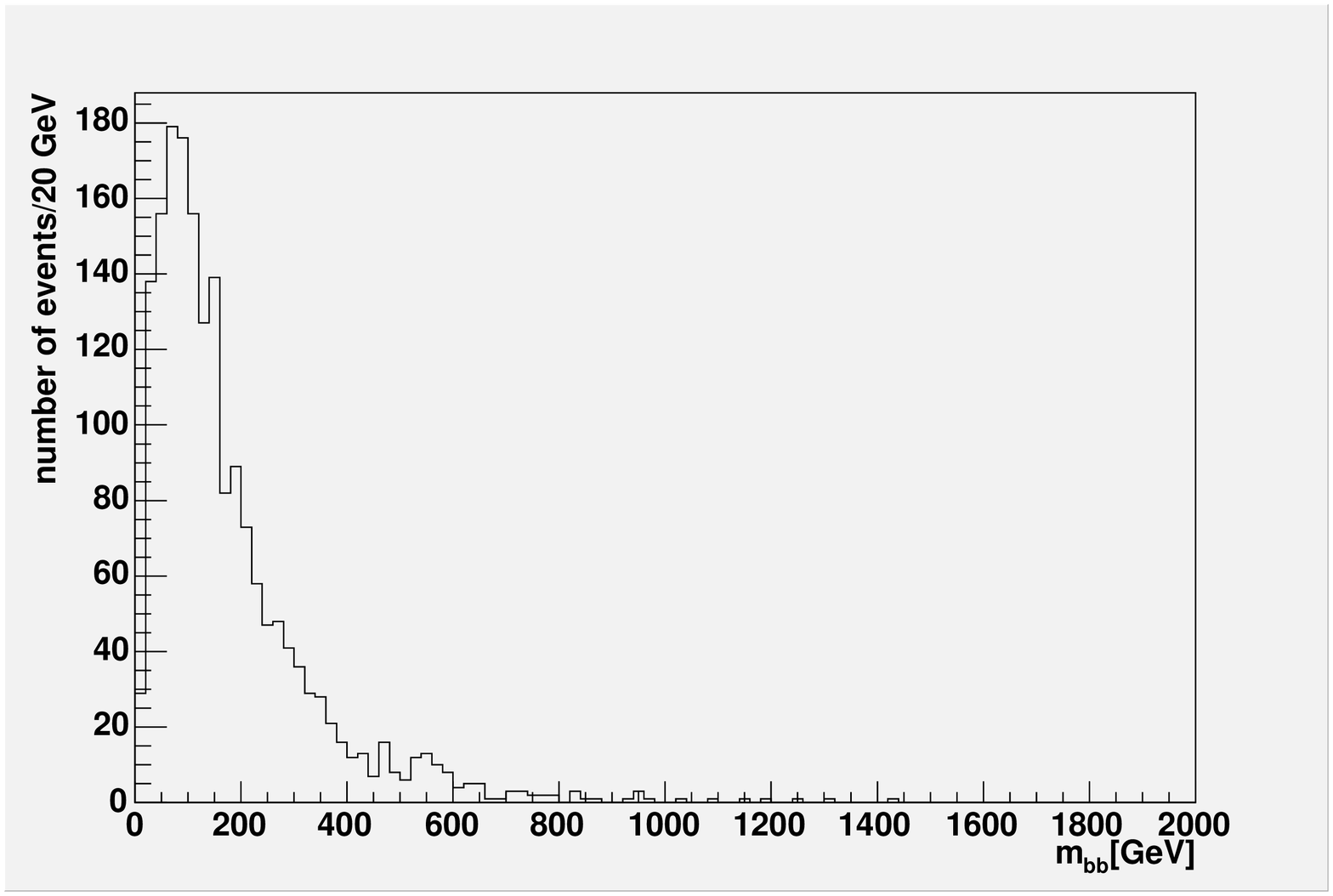,height=90mm,width=55mm,clip=,silent=,angle=-90,
bbllx=2.5cm,bblly=1cm,bburx=20cm,bbury=30cm}
\put(-2.4,-1.1){(b)}
\put(-4.5,-2.1){\small N=1819}
\end{picture}}
\put(-0.4,11.4){\begin{picture}(9,5.5)
\epsfig{file=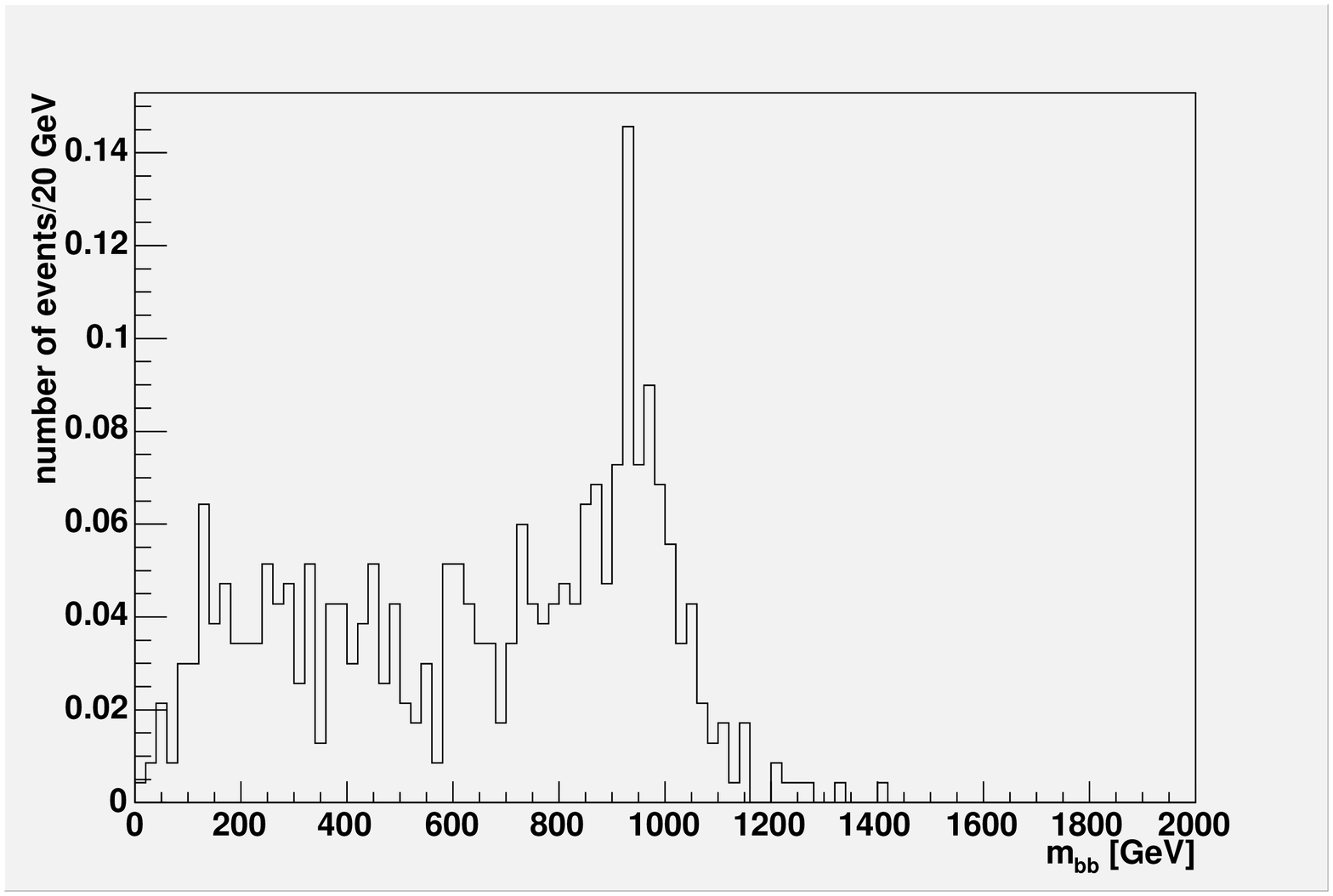,height=90mm,width=55mm,clip=,silent=,angle=-90,
bbllx=2.5cm,bblly=1cm,bburx=20cm,bbury=30cm}
\put(-2.4,-1.1){(a)}
\put(-4.5,-2.1){\small N=2.34}
\end{picture}}
\end{picture}

\vspace{-2mm} 

\noindent Fig. 2. Reconstructed
distributions of the invariant mass $m_{\mathrm{bb}}$
of the b$\mathrm{\bar{b}}$ pair obtained for the 
$\rho \rightarrow$ b$\bar{\mathrm{b}}$
process with $M_{\rho}$=1000~GeV
and $\Gamma_{\rho}=42.4$~GeV (a), and for the QCD background (b).

\bigskip

\noindent  ACKNOWLEDGEMENT: We wish to thank Jonathan Ferland, University
of Montreal, for his help with the reconstruction procedure.

\end{document}